\documentclass[letter,12pt]{article}
\usepackage[margin=35mm]{geometry}
\usepackage{multirow}
\usepackage{adjustbox}
\usepackage{amsmath}
\usepackage{amsfonts}
\usepackage{amssymb}
\usepackage{graphicx}
\usepackage{verbatim}
\immediate\write18{texcount -tex -sum  \jobname.tex > \jobname.wordcount.tex}

\providecommand{\keywords}[1]
{
  \small	
  \textbf{\textit{Keywords---}} #1
}

\title{Recent Advances in Software Effort Estimation using Machine Learning}
\author{V\'ictor Uc-Cetina \\
\small uccetina@correo.uady.mx \\
\small Facultad de Matem\'aticas \\
\small Universidad Aut\'onoma de Yucat\'an \\
}

\begin{document}

\maketitle

\begin{abstract}
An increasing number of software companies have already realized the importance of storing project-related data as valuable sources of information for training prediction models. Such kind of modeling opens the door for the implementation of tailored strategies to increase the accuracy in effort estimation of whole teams of engineers. In this article we review the most recent machine learning approaches used to estimate software development efforts for both, non-agile and agile methodologies. We analyze the benefits of adopting an agile methodology in terms of effort estimation possibilities, such as the modeling of programming patterns and misestimation patterns by individual engineers. We conclude with an analysis of current and future trends, regarding software effort estimation through data-driven predictive models.
\end{abstract}

\\
\maketitle

A major goal of software project managers and software developers is to calculate accurate estimates of the effort required to complete a software development project. Effort prediction in software engineering projects is considered by some software development managers as a mission impossible kind of task. Some other managers think it is certainly possible to accomplish it, at least for specific categories of software applications, if enough historical information is available and the right methodology is chosen \cite{Kitchenham2002}. The common scenario is that each software company ends up developing its own strategy over time, depending on the kind of applications they develop. In this article we review the most recently proposed machine learning methods for software effort prediction, and we categorize them in non-agile and agile methods.

It is well known that machine learning (ML) has become a very robust data-driven tool for solving a variety of problems, ranging from face recognition to chatbots development, from product recommendation to protein structure prediction. We have been noticing significant impact in different fields such as computer vision, natural language processing, e-commerce, bioinformatics, robotics, audio processing, etc. In general applications and fields benefiting from ML algorithms have been multiplying in the last decade.  The same can be said about ML applications for software engineering problems. In a recent survey \cite{Yang2021}, five software engineering activities were clearly identified as the most investigated using deep ML, namely,

\begin{itemize}
\item Software design, 
\item Software implementation, 
\item Software testing and debugging, 
\item Software maintenance, and 
\item Software management.
\end{itemize}

Each of these activities include a different number of more specific research topics. For instance, in software management, two are the most common topics:

\begin{itemize}
\item Software repository mining, and 
\item Effort cost prediction.
\end{itemize}

Predicting projects efforts is by itself a difficult task. Estimation or prediction of software development effort is not an exception. In order to estimate with the highest of prediction ability we need to exploit past information, we need to develop data-driven strategies. As Jørgensen argues \cite{Jorgensen2007}, historical data relevant to the target project improves estimation accuracy. Moreover, based on recently published well-supported research results \cite{Jorgensen2014}, he says that, what we currently know about software effort and cost estimation is not enough to solve the estimation challenges in the software industry. Nevertheless, it helps us to identify several actions that might improve estimation accuracy.

In this article we are interested in the effort prediction problem for whole projects or large modules of software (non-agile prediction), and also in agile software effort prediction.

\section{NON-AGILE EFFORT PREDICTION}

In one of the main works on effort prediction carried out two decades ago Kitchenham et al. \cite{Kitchenham2002} studied the effort and duration of 147 projects, including maintenance and development ones. This data set was obtained from one software company of outsourcing. They discovered that 63\% of the estimates were calculated within 25\% of the actual value, with an average absolute error of 0.26. In other words, this study supports the idea that estimation of software development efforts is achievable with a reasonable margin of error, small enough to consider it as a useful practice in many cases.

One of the earliest studies that used machine learning for building estimators of software development effort from historical data was presented by Srinivasan et al. \cite{Srinivasan1995}. In this work, regression trees and multilayer perceptrons algorithms were compared against traditional parametric models for  effort prediction. In both cases the results show that machine learning algorithms accomplished competitive performances. One of the main advantages of the methods studied is their adaptability to new data sets. 

Jørgensen et al. \cite{Jorgensen2007} systematically reviewed 304 software cost estimation articles, classifying them according to research topic, estimation approach, research approach, study context, and use of data sets. Based on this systematic categorization of papers, the authors make some recommendations over changes they identified as being useful in estimation research.

There are some data-driven models typically studied to solve the effort prediction software. Two categories that have been compared are linear regression models and neural networks, being the second type more complex models with many more parameters to be estimated. Lopez-Martin et al.  \cite{Lopez-Martin2015} performed a comparison of these two categories of algorithms. Specifically they compared a multiple linear regression methods with multilayer perceptrons and radial basis function networks. Their conclusion is somewhat expected: neural networks showed higher accuracy than multiple linear regression.

In a similar approach Bisi et al. \cite{Bisi2016} tried a methodology based on evolutionary computation combining swarm optimization for training, principal component analysis to reduce the dimension of the input data, and finally genetic algorithms to optimize the weights of a neural network, also reporting competitive performances.

More recently, BaniMustafa et al. \cite{BaniMustafa2018} evaluated 3 machine learning models using the COCOMO NASA data set which includes 93 software projects. The algorithms used in this study are naïve Bayes, logistic regression, and random forests. Although these are not by any means state-of-the-art algorithms, the approach is of general applicability and provide a baseline of methods to start with and further increase the robustness of our methodology for effort prediction. The data set includes 24 attributes such as the category of the application (with 14 possibilities, including avionics, communications, science, and simulation), development mode (indicating one of the following options: embedded, organic, semidetached), lines of code measure and the actual effort in months. The experimental results show a competitive performance of all 3 methods over the COCOMO model, being the random forest model the best one. This is not surprising considering that random forest is a much more robust method compared to the two other tested algorithms.

Effort predictions based on team size has been also approached by Rai et al. \cite{Rai2021}. Recently, they proposed the use of the constructive cost model known as COCOMO \cite{Boehm81} together with support vector regression, to predict effort estimation.

\section{AGILE EFFORT PREDICTION}

Effort estimation in agile projects focuses on estimating the effort required for completing user stories. In a recent article, Choetkiertikul et al. \cite{Choetkiertikul2019} proposed a prediction model for estimating story points based on the combination of long short-term memory and recurrent highway networks. Their empirical evaluation with 16 open source projects demonstrates that this approach consistently outperforms baselines such as random guessing, mean, and median methods and six other more complex methods such as Doc2Vec and random forests. Also, Ochodek et al. \cite{Ochodek2020} investigated several neural networks testing them with 437 use cases from 27 software development projects. The main goal of this study was to develop an easy-to-train method, which is comparable in performance to the existing methodologies. The experimental results shows as the best performer a convolutional neural network used together with a word-embedding model.

Phannachitta et al. \cite{Phannachitta2020} carried out a systematic comparison of software effort estimators on 13 standard benchmark datasets. Performance metrics and statistical test methods were used together with different machine learning algorithms. The experimental results show that an analogy-based model that is adapted through a gradient boosting machine algorithm and a traditional adaptation method based on productivity adjustment is the best performer.

Sarro et al. \cite{Sarro2020} introduced the approach called learning from mistakes. The main idea is focused on automatically learning from past estimation errors made by human experts. These errors are used to improve the accuracy of the predictions by calculating possible future errors in estimates. This empirical investigation included 402 maintenance and development projects. The main finding of this study is that the type, severity and magnitude of errors are all predictable. The algorithms used in this work to predict errors are classification and regression trees, k-nearest neighbors, naïve Bayes, and linear programming.

Tawosi et al. \cite{Tawosi2021} replicated and extended a study \cite{Sarro2016} on multi-objective software effort estimation in order to increase the confidence in those previous results. Later on, Tawosi et al. \cite{Tawosi2022b} carried out a replication study of the analysis previously presented by Ochodek et al. \cite{Ochodek2020}. This time the authors used a larger dataset. However, the results they obtained are not supporting the hypothesis stating that deep neural networks can significantly outperform less sophisticated methods for predicting actual efforts in software development. Moreover, they concluded that semantic analysis of the texts describing the stories is not enough and it is prone to introduce errors in the estimations.

There are a couple of datasets usually employed for researching effort prediction in agile software development projects: the Choetkiertikul dataset \cite{Choetkiertikul2019} and the TAWOS dataset \cite{Tawosi2022}.

The Choetkiertikul dataset was created in 2019 as part of a study that investigated the use of a deep learning model for estimating story points. This dataset is stored in a csv format file and contains 23,313 issues from 16 open-source projects. The data can be mainly used for story points-based estimation. 

On the other hand, the TAWOS dataset is a dataset created in 2022 from Jira repositories and it includes 44 agile open-source software projects with a total of more than 500,000 issues. It is an easy-to-use dataset stored as a relational database that can be used with MySql workbench for investigating different software engineering problems, i. e. agile software effort estimation. For example, Tawosi et al. \cite{Tawosi2022c} used that dataset to investigate how clustering could be used to estimate story points in agile software development projects.

\section{NON-AGILE VS. AGILE}

\begin{figure}[h]
\centering
\includegraphics[width=5cm]{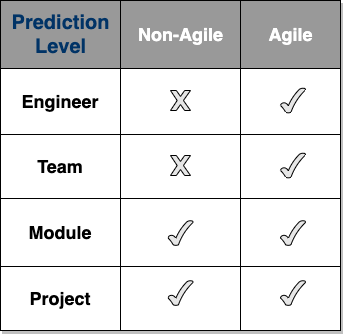}
\caption{Benefits of performing effort prediction under an agile software development methodology.}
\label{Agile}
\end{figure}

Effort estimation in non-agile project is usually carried out at the level of the whole project or whole modules, which increases the prediction error, simply due to the amount of lines of codes that need to be considered. When the project is developed using an agile methodology, the prediction can be performed at the level of individual engineers, passing through teams-level, up to module and project-level, as summarized in Figure \ref{Agile}.

One way of achieving more precise effort estimation in agile development projects is by keeping track of stories effort estimations and stories actual efforts needed by engineers of software teams. We argue that when this information is stored, the following benefits are available:

\begin{enumerate}
\item Calculation of the current misestimation rate of each engineer.
\item Possibility of giving data-driven feedback when new estimations are needed.
\item Identification of those engineers with above average skills in you team or company, for specific tasks or technologies.
\end{enumerate}

Moreover, we believe that:

\begin{enumerate}
\item As software engineers become more experienced in designing stories, their stories gradually become more compact and are more easily attainable. We consider this as a highly desirable design skill.

\item As a result of being able to design more compact stories, stories become more predictable in terms of needed effort and therefore software engineers gradually reduce their effort misestimation.
\end{enumerate}

Consequently, it is reasonable to expect that over the years, the percentage of misestimated stories by an engineer will decrease. In other words, there are patterns inherent to the process that an engineer follows to design stories and estimate their complexity and those patterns can be learned using machine learning models.

\section{CURRENT AND FUTURE TRENDS}

We can identify four main families of methods used in recent approaches for competitively predicting software development effort, all of them with applications in non-agile and agile software effort prediction (See Figure \ref{predictive_methods}):

\begin{figure*}
\centering
\includegraphics[width=15cm]{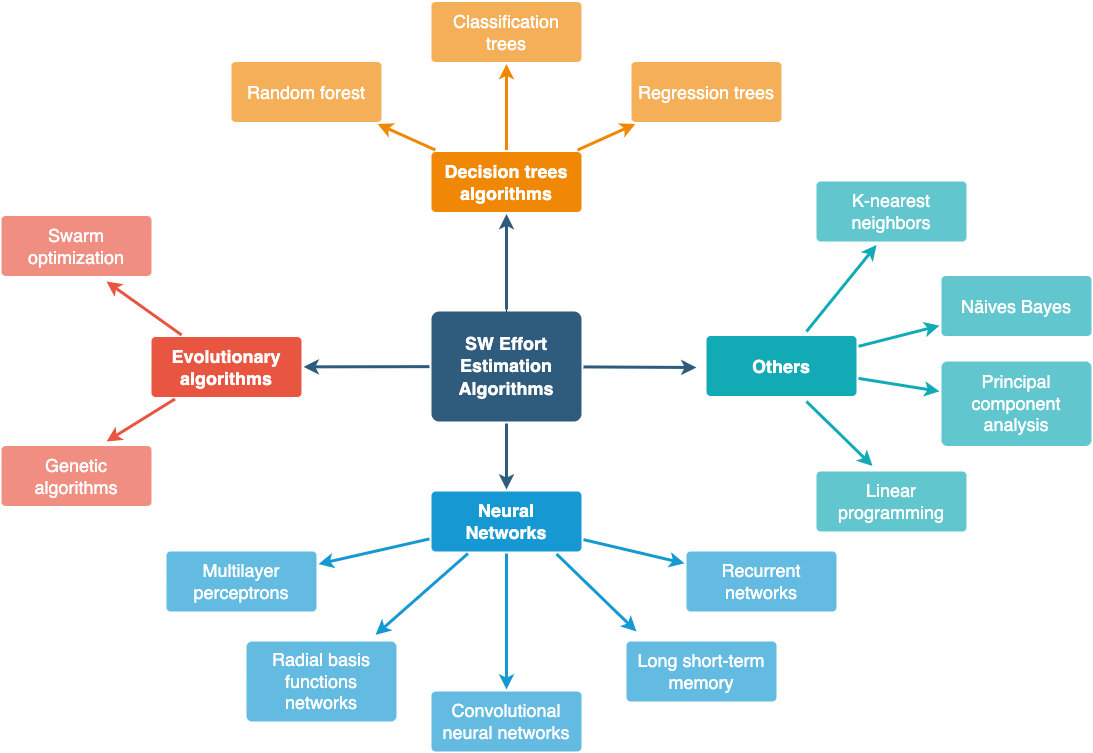}
\caption{Four main families of methods recently used for software effort prediction in non-agile and agile methodologies.}
\label{predictive_methods}
\end{figure*}

\begin{itemize}
\item Evolutionary algorithms,
\item Decision trees algorithms,
\item Shallow and deep neural networks, and
\item Others.
\end{itemize}
 
On the other hand the predictions in all these approaches are calculated differently, depending on the historical information available. Some of the features used as inputs for the predictive models are:

\begin{itemize}
\item Attributes describing the complexity (e.g. lines of codes),
\item Attributes describing the type of software application (e.g. mobile app),
\item Names of use cases,
\item Features extracted using analogy-based methods,
\item Time required to accomplish stories, and
\item Complexity points of user stories.
\end{itemize}

In spite of the fact that deep neural networks are currently a dominating force in practically all domains where data-driven predictions are needed, we argue that in the particular case of effort prediction of agile software stories, even simple methods such as linear regression models can be very effective and computationally more suitable to be implemented and maintained in software development management systems. 

However, we expect to see more neural network models being applied to the effort prediction problem. For instance, transformer architectures \cite{Vaswani2017}, based on attention layers, have revolutionized the language processing domain. They are currently being investigated for different tasks in the computer vision domain, and we will not have to wait much before transformers start to be applied as predictive models in software engineering tasks, such as effort prediction.

In recent years, more and more companies have started using various types of project management software to keep track of their daily development tasks and as a result of this practice, larger data sets are being collected. Those data sets become over time an important asset, allowing companies to generate more accurate prediction models in the future.

From our point of view, in the coming years we will witness an increment in tracking and management software systems with a diversity of data-driven prediction features. The prediction of efforts at different levels of complexity, ranging from individual software stories up to whole modules or projects will definitely be one of these features.

\section{CONCLUSION}
Effort prediction of software development is attainable with some degree of accuracy, making it a viable practice in industry. Modern neural network models are capable to learn from historical data of software development projects and extract critical patterns that allow them to predict efforts in similar projects. 

However, when the project to be analyzed for effort prediction is particularly different to the projects in our available historical data, then the accuracy in prediction decreases significantly. Therefore more research is still needed to solve this kind of extrapolation in the estimation of project efforts. 

Finally, since an increasing number of software companies are adopting an agile software development methodology, effort estimation methods focusing on tracking individual engineers development patterns will be developed.

\bibliographystyle{abbrv}
\bibliography{references.bib}

\end{document}